\documentclass[groupedaddress,superscriptaddress,amsmath,amssymb,aps,prl,
twocolumn,10pt,floatfix,longbibliography,footnoteinbib]{revtex4-2}

\usepackage{amsmath,amssymb,amsfonts,bbm,color,graphicx}

\usepackage[utf8]{inputenc}
\usepackage[T1]{fontenc}

\usepackage[unicode=true,colorlinks=true]{hyperref}
\hypersetup{linkcolor=blue,citecolor=blue,urlcolor=blue}

\usepackage{bm}
\usepackage{dsfont}

\usepackage[normalem]{ulem}

\usepackage{wasysym}
\usepackage{braket}

\usepackage{slashed}

\newcommand{\iu}{\mathrm{i}} 
\newcommand{\eu}{\mathrm{e}} 
\newcommand{\du}{\mathrm{d}} 

\newcommand{\bvec}[1]{{\bm{#1}}}

\newcommand{\SOtwo}{\mathrm{SO(2)}}

\newcommand{\ph}{\mathrm{\ph}}

\usepackage[svgnames]{xcolor}

\begin{document}


\title{Anisotropic vacancy-induced magnetization textures in altermagnets}

\author{Ruben Burkard}
\affiliation{Institut für Theoretische Physik, Universit\"at Köln, Zülpicher Straße 77a, 50937 Köln, Germany}

\author{Mathias S. Scheurer}
\affiliation{Institute for Theoretical Physics III, University of Stuttgart, 70550 Stuttgart, Germany}

\author{Urban F. P. Seifert}
\affiliation{Institut für Theoretische Physik, Universit\"at Köln, Zülpicher Straße 77a, 50937 Köln, Germany}


\begin{abstract}
We study magnetic textures induced by vacancies in altermagnets using microscopic simulations and low-energy field theory.
We show that a vacancy generically produces a real-space anisotropic distortion of the magnetic order, whose structure encodes the symmetry of the underlying altermagnetic state. This impurity response offers a direct route to detecting altermagnetic order with locally resolved probes. We demonstrate this for both classical altermagnets, where vacancies generate anisotropic magnetization textures in a transverse magnetic field, and quantum models, where fluctuations induce longitudinal power-law decaying magnetic distortions even at zero field.
\end{abstract}

\date{\today}

\maketitle


\textit{Introduction.---}%
Altermagnetism has emerged as a unifying theme for unconventional magnets featuring time-reversal-symmetry broken spectra and response functions while the net magnetization vanishes by symmetry \cite{libor22a,libor22b}.
They not only hold promise for future spintronics applications but the interplay of magnetic, orbital and electronic degrees of freedom has also attracted strong interest into altermagnets as strongly-correlated systems~\cite{Fukaya2025Aug,Jungwirth2025Aug}.

Key to the definition of an altermagnetic phase is a symmetry of joint spin rotations/time-reversal symmetry and crystal symmetry operations which relate magnetic sublattices other than translation and/or inversion.
As a result, spatial multipoles of the uniform magnetization (spin density) become finite and act as secondary order parameters distinguishing altermagnets from conventional antiferromagnets \cite{fernandes24,mcclarty24,bhowal24}. In itinerant systems, this leads to momentum-dependent spin splittings, e.g. with a $d$-wave form factor, in the electronic spectrum \cite{hayami19,gonzalez21,libor22a,libor22b,roig2024}.

Significant theoretical and experimental progress has been made in identifying signatures of altermagnetism, foremost spectroscopic \cite{maier23,smejkal23,liu24,bittencourt26,pupim26} and transport probes \cite{gonzalez21,libor22,fang24,leeb24,lin25,takahashi25}, which are sensitive to the spin-dependent spatial anisotropy of \emph{excitations} (electrons or magnons) in the altermagnetic state.
In contrast, in this Letter, we show that in altermagnets, non-magnetic impurities lead to distortions in the ordering of local moments which directly resolve the \emph{altermagnetic order itself} and could be probed using nanoscale techniques such as spin-polarized scanning tunneling spectroscopy \cite{wiesendanger09}, NV magnetometry \cite{rovny24} or x-ray microscopy methods, which have recently been used to image order parameter defects in altermagnets \cite{amin24,yamamoto25}.
Indeed, probing the response to impurities is a powerful method to investigate a system's ground state and its correlations \cite{alloul09}.
This includes vacancy-induced staggered magnetizations near vacancies in quantum antiferromagnets \cite{julien00,bobroff09}, universal scaling forms for impurity-induced textures and dynamics \cite{sachdev99, sachdev03,metlitski07} and long-ranged distortions of non-collinear magnetic order \cite{wollny11,consoli24,zhitomirsky25}.

In this context, the intertwinement of magnetic ordering with the crystalline and orbital environment renders altermagnets particularly susceptible to crystalline impurities as symmetry-breaking defects. Since such impurities are experimentally inevitable, understanding the impurity response of altermagnets is a timely question, both as a stepping stone for assessing disorder effects but especially also for exploring them as local probes.

Previous works on inhomogeneities in altermagnets have been mostly concerned with the impact on electronic quasiparticles \cite{chen24,maiani25,sukhachov24,QPIAgterberg,2025arXiv251018102B,2025arXiv251019943S,PhysRevB.111.035132,PhysRevB.111.L100502,2026arXiv260214950M,schrade26}, see also recent experimental works \cite{fu25,gu26,mu26,yang26}, rather than the response of the local magnetic moments themselves, which is the subject of this Letter.

\begin{figure}[t]
    \centering
    \includegraphics[width=\columnwidth]{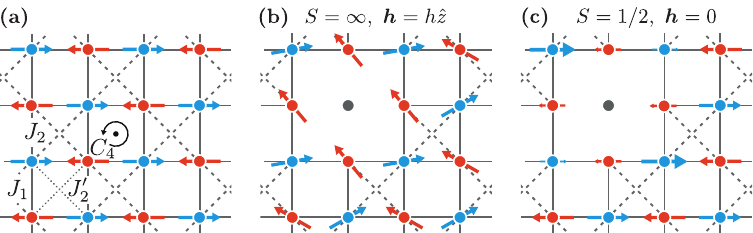}
    \caption{(a) Illustration of checkerboard lattice Heisenberg model as a minimal model for an altermagnet. For visual clarity, we omit showing the dotted lines representing $J_2'$ in all but one plaquette. (b) In the classical model (i.e.~in the large-$S$ limit), a vacancy leads to distortions in the local moment's canting towards an applied magnetic field $\bvec{h}$, resulting in a magnetization texture \emph{transverse} to the ordering axis. (c) In quantum altermagnets, even at zero field, a vacancy induces inhomogeneities in fluctuation-induced local moment renormalizations, leading to \emph{longitudinal} magnetization textures.} 
    \label{fig:illustration-lattices}
\end{figure}

We distinguish ``classical'' altermagnets, corresponding to the large-spin limit of spin-$S$ magnets, from ``quantum'' altermagnets with $S<\infty$, see also Fig.~\ref{fig:illustration-lattices} for a schematic overview.
In the classical limit, a vacancy preserves the antiparallel alignment of local moments in a collinear altermagnet, but we show that in an applied magnetic field, a spatially anisotropic magnetization texture near the vacancy emerges.
This texture inherits the symmetry of the altermagnetic order parameter, and we envision that this enables real-space imaging of the multipolar form factor in an altermagnet.
In quantum altermagnets, a non-magnetic impurity produces anisotropic distortions already at zero field, both in the local longitudinal staggered magnetization and in the slowly decaying vacancy-induced longitudinal magnetization where we show that altermagnetism leads to a unique multipolar contribution.


\textit{Symmetry considerations.---}%
An ideal altermagnet (in the limit of zero spin-orbit coupling) is a state of collinear local moments where the magnetic sublattices are related by some point-group operation $g$, but \emph{not} translation or inversion \cite{libor22a,mcclarty24}. Thus, $\Theta g$ is a symmetry of the altermagnet, with $\Theta$ denoting time-reversal.
As a vacancy on one magnetic sublattice breaks $\Theta g$, we deduce that (i) a finite magnetization near the vacancy becomes symmetry-allowed, and (ii) vacancy-induced distortions of the magnetic order will generally be anisotropic: by definition, $g$ is not in the site-symmetry group and thus a single vacancy can not be a symmetric perturbation.
We contrast this with conventional antiferromagnetic order, which allows scenarios where a vacancy can be placed on a magnetic site which is invariant under all point group operations; then, the resulting distortions are fully symmetric, as is the case for the square lattice antiferromagnet~\cite{vojta00, eggert07}.

One may characterize altermagnets in terms of a Landau theory for the staggered magnetization $\bvec{N}$ and the uniform magnetization $\bvec{M}$ which transform in a non-trivial ($\Gamma_N$) and the trivial ($\Gamma_1$) irreducible representation (IR) of the point group, respectively \cite{steward23,fernandes24,bhowal24,mcclarty24}.
Crucially, in the presence of a vacancy, the system is no longer homogeneous and symmetry-allowed contributions to the free energy of the form
\begin{equation} \label{eq:f-am}
	F_\mathrm{AM}[\bvec{N},\bvec{M}] \sim \bvec{M} \cdot \mathcal{D}[\partial_x,\partial_y,\partial_z] \bvec{N},
\end{equation}
become important.
Here, $\mathcal{D}[\partial_x,\partial_y,\partial_z]$ is a derivative operator that transforms in the same IR of the point group as $\bvec{N}$, for example $\mathcal{D}[\partial_x,\partial_y] = 2 \partial_x \partial_y$ for $d_{xy}$-wave altermagnets where $\bvec{N}$ transforms, for example, in the $B_2$ IR of the $C_{4v}$ point group.
Hence, spatial variations of $\bvec{N}$ can induce non-trivial magnetization textures, $\bvec{M}(\bvec{r}) \sim \mathcal{D}[\partial_x,\partial_y,\partial_z] \bvec{N}(\bvec{r})$, as also observed in altermagnetic domain walls \cite{gomonay24,schrade26}.


\textit{Vacancy in classical altermagnets: magnetization textures.---}%
We first consider ``classical altermagnets'' where the local moments are described in terms of classical vectors with length $\bvec{S}_i^2 = S^2$, interacting via the Heisenberg Hamiltonian
\begin{equation} \label{eq:h-j}
	H = \sum_{ij} J_{ij} \bvec{S}_i \cdot \bvec{S}_j.
\end{equation}
Here $J_{ij}$ are exchange couplings stabilizing collinear compensated magnetic order $\bvec{S}_i = \eta_i S \hat{n}$ along an axis $\hat{n}$ and $\eta_i = \pm 1$ on the two magnetic sublattices.
The $\SOtwo$ symmetry of spin rotations around $\hat{n}$ is preserved by the vacancy.
Thus, no transverse distortions emerge, and since at zero temperature ($T=0$) there are no fluctuations to renormalize the ordered moment, a vacancy cannot distort classical collinear magnetic order.

Now, consider adding a magnetic field $\bvec{h}$ via $H\to H - \bvec{h} \cdot \sum_i \bvec{S}_i$.
Then, the local moments cant towards the field direction to produce a finite uniform magnetization, which is maximized when $\bvec{h} \perp \bvec{N}$.
Crucially, near a vacancy, this canting angle can vary from its bulk (clean) value, inducing a finite component of the staggered magnetization $\bvec{N}$ \emph{along} the field axis \cite{eggert07}.
In altermagnets, given above symmetry considerations, this field-induced vacancy response is necessarily anisotropic.

We make these considerations explicit by numerical simulations of the checkerboard lattice Heisenberg model $H_\mathrm{CB}$ as a prototypical altermagnetic model \cite{kaushal25}, relevant to quasi-2D oxychalcogenides such as KV$_2$Se$_2$O \cite{jiang25}.
The coupling constants are given by $J_{i,i+\hat{x}} = J_{i,i+\hat{y}} = J_1$, $J_{i,i+\hat{x}+(-1)^i\hat{y}} = J_2$ and $J_{i,i+\hat{x}-(-1)^i\hat{y}}= J_2'$ with $\hat{x}(\hat{y})$ denoting unit vectors in $x(y)$-direction and $(-1)^i :=( -1)^{i_x+i_y}$, see also Fig.~\ref{fig:illustration-lattices}(a).
For $J_1 > |J_2|,|J_2'|, J_2+J_2', 1/8h$, the ground state exhibits collinear Néel order.
Crucially, for $J_2 \neq J_2'$, the two magnetic sublattices are neither related by lattice translations nor by inversion.
Instead, the magnetic order preserves the combination of $C_4$ lattice rotations (around a plaquette center) and time reversal $\Theta$, rendering it a $d$-wave altermagnet.

\begin{figure}[t]
    \includegraphics[width=\linewidth]{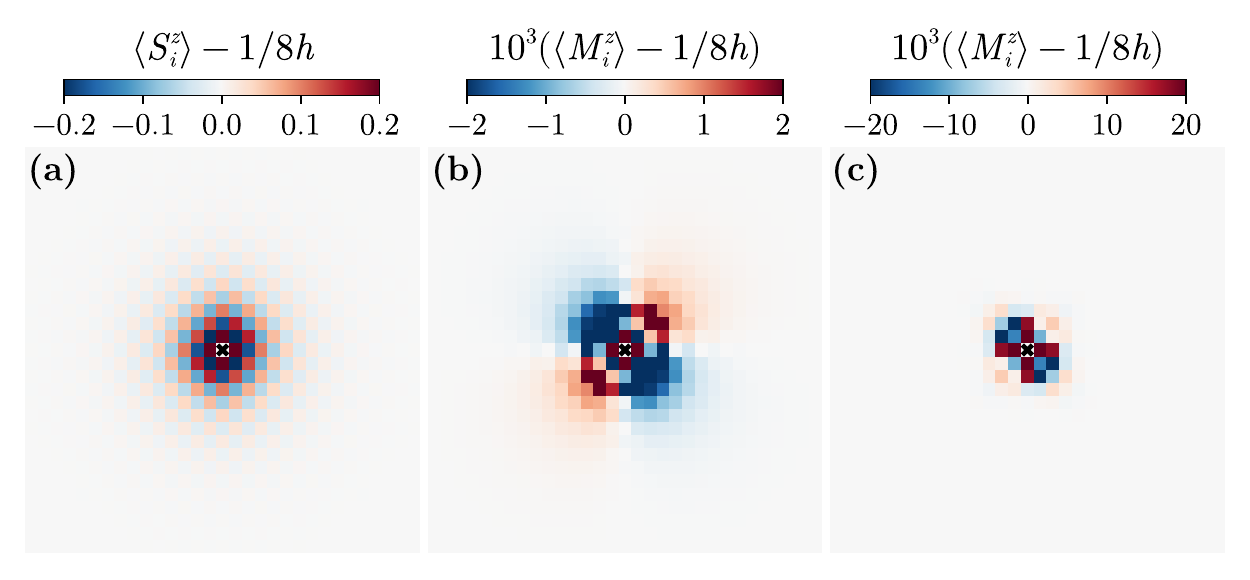}
    \caption{The classical Heisenberg model with altermagnetic interactions and a single vacancy in a small perpendicular magnetic field. (a) Site-resolved spin component $S_i^z$ (along the field direction $\bvec{h} = h \hat{z}$) for $J_2=0.6J_1$, $J_2'=0$, and $h=0.6J_1$. The position of the vacancy is marked with an 'x'. The local magnetization around the vacancy is shown in (b) for the same parameters as in (a), and in (c) for $h=4 J_1$ and $J_2=0.6 J_1, J_2'=0$.}
    \label{fig:results_classical}
\end{figure}

The classical spin configuration of the model with vacancy at site $\bvec{r}=(0,0)$ and in an applied field $\bvec{h} = h \hat{z}$ is obtained via iterative relaxation.
The vacancy-induced texture can be characterized via the coarse-grained magnetization field $\bvec{M}(\bvec{r})$, which we obtain from the microscopic spin configuration via a Gaussian filter $\bvec{M}(\bvec{r}) = \sum_{i,j=-1}^1 \omega_{ij}\bvec{S}_{\bvec{r}+i \hat{x}+j \hat{y}}$ with the kernel $\omega_{ij} = (v v^\top)_{ij}$ where $v = (1,2,1)^\top / 4$.
This yields a \emph{local} average of the uniform magnetization, experimentally accessible using spatially-resolved probes \cite{wiesendanger09,rovny24,amin24}.

Results for the out-of-plane spin component $S^z_i$ in the ground state are shown in Fig.~\ref{fig:results_classical}(a): the vacancy induces a finite staggered magnetization along the field direction, on top of the uniform magnetization due to canting, as illustrated in Fig.~\ref{fig:illustration-lattices}(b).
The vacancy-induced texture in the local magnetization $\bvec{M}(\bvec{r})$, shown for two different field strengths in Fig.~\ref{fig:results_classical}(b) and Fig.~\ref{fig:results_classical}(c), is found to be strongly anisotropic.
Crucially, for sufficiently small applied magnetic fields, the texture in $\bvec{M}(\bvec{r})$ is approximately odd under $C_4$ lattice rotations and exhibits an approximate $d_{xy}$-wave symmetry.
We obtain similar results for a vacancy in an honeycomb-lattice altermagnet, see also Ref.~\onlinecite{suppmat}.


\textit{Continuum field theory.---}%
To elucidate the observed vacancy-induced magnetization textures, we turn to a low-energy field theory.
We decompose the spins into uniform and staggered magnetization fields as
\begin{equation} \label{eq:s-n-m}
	\bvec{S}_i = (-1)^{i} \bvec{N}(\bvec{r}_i) \sqrt{S^2-\bvec{M}^2(\bvec{r}_i)}+\bvec{M}(\bvec{r}_i),
\end{equation}
where the unit-length constraint is satisfied if we demand $|\bvec{N}|^2=1$ and $\bvec{N} \cdot \bvec{M} = 0$.
Inserting this into the microscopic Hamiltonian, expanding in small $|\bvec{M}| \ll S$ and taking the continuum limit, we arrive at the Hamiltonian density $\mathcal{H} = \mathcal{H}_\mathrm{bulk} + \mathcal{H}_\mathrm{vac}$, 
\begin{subequations}\begin{align}
        \mathcal{H}_\mathrm{bulk} &= \frac{1}{  2 \chi_\perp} \bvec{M}^2 + S^2\frac{\rho_s}{2} (\nabla \bvec{N})^2 + SJ_A \bvec{M} \cdot 2\partial_x \partial_y \bvec{N}  \notag\\
        &\qquad- \bvec{h}\cdot\bvec{M}+\lambda \bvec{M} \cdot  \bvec{N}+ \mu (\bvec{N}^2-1) \label{eq:h-bulk} \\
    \mathcal{H}_\mathrm{vac} &= \delta(\bvec{r}) f(\bvec{M},\bvec{N}) \label{eq:h-vac}
\end{align}\end{subequations}
with the inverse spin susceptibility $\chi_\perp^{-1}= 8J_1$, stiffness $\rho_s=J_1-J_2-J_2'$ and effective altermagnetic coupling $J_A= J_2-J_2'$ between uniform and staggered magnetization as anticipated in Eq.~\eqref{eq:f-am}.
Here, $f(\bvec{M},\bvec{N})$ includes the same interaction terms as $\mathcal{H}_\mathrm{bulk}$ with different bare coupling constants and also new terms due to the broken rotational symmetry around the center of a square, see Supplemental Material for an explicit expression \cite{suppmat}.
Although the \emph{classical} bulk theory at zero temperature $T=0$ does not exhibit fluctuation-induced renormalizations of its couplings, a vacancy acts as a UV defect and thus, at low energies, the effective vacancy couplings will flow away from their bare (lattice-scale) values.
However, the leading-order long-distance behavior of $\bvec{M}(\bvec{r})$ as derived below is cutoff-independent and we may therefore continue with the bare coupling constants in $f(\bvec{M},\bvec{N})$.

Next, via the saddle-point equations, we obtain the magnetization away from the vacancy as
\begin{multline} \label{eq:mag-classical}
     \bvec{M} = \chi_\perp \Bigl[ \left(\bvec{h}- \bvec{N} (\bvec{h}\cdot \bvec{N}) \right)
     \\
     -S J_A \left( 2\partial_x \partial_y \bvec{N}+ \bvec{N}(\bvec{N} \cdot 2\partial_x \partial_y \bvec{N}  ) \right)\Bigr],
\end{multline}
comprised of a uniform response to an applied field $\bvec{h}$ and contributions induced by gradients in $\bvec{N}$ via the altermagnetic coupling.
Proceeding, we use Eq.~\eqref{eq:mag-classical} to integrate out $\bvec{M}(\bvec{r})$ from $\mathcal{H}_\mathrm{bulk}$ in Eq.~\eqref{eq:h-bulk}, and similarly in Eq.~\eqref{eq:h-vac} \cite{suppmat}.
Without loss of generality, we consider the staggered magnetization to lie along the $\hat{x}$-axis and $\bvec{h} = h \hat{z}$.
We parametrize $\bvec{N}$ in terms of fluctuations around the ordered state, i.e., 
\begin{equation} \label{eq:condensed_Neel_expansion}
    \bvec{N} = \left( \sqrt{1- n_y^2- n_z^2}, \, n_y, \, n_z \right),  \quad n_{y,z} \ll1,
\end{equation}
which, upon inserting into $\mathcal{H}_\mathrm{bulk}$ and expanding to quadratic order, yields two decoupled Gaussian theories, for the linearly dispersing Goldstone mode $n^y$, and for $n^z$ which becomes gapped due to $h\neq 0$.
Proceeding analogously for $\mathcal{H}_\mathrm{vac}$, we find up to linear order in $h$ four separate source terms for the massive mode, $\mathcal{H}_\mathrm{vac}[n_z]\! =\!\!  \delta(\bvec{r}) \! \! \left[Sh n_z \!-\!S\chi_\perp \!J_A h 2\partial_x \partial_y n_z \!+\!S^2\! J_A   n_z 2\partial_x \partial_y n_z  \!\!+\! \!S^2\!\rho_s n_z  \Delta n_z \right]$.
We obtain the staggered magnetization along the field direction $n_z(\bvec{r})$ as the linear response of the bulk theory to the defect term $\mathcal{H}_\mathrm{vac}[n_z]$ (controlled for small fields $h$ and at large distances away from the vacancy, where gradients are small),
\begin{align} \label{eq:nz-classical}
    \langle n_z(\bvec{r})\rangle \approx \frac{h}{2\pi \rho_s S} K_0 \Bigl(\frac{h}{cS} r  \Bigr) \Bigl[ 1+\mathcal{O}(h^2)\Bigr],
\end{align}
where we write $\bvec{r} = (r \cos \theta, r \sin \theta)^\top$ in polar coordinates. The first term in square brackets also emerges in conventional antiferromagnets \cite{eggert07}, and the $\mathcal{O}(h^2)$ term depends on the UV-cutoff.
Using this in Eq.~\eqref{eq:mag-classical} we arrive at our key result for the out-of-plane magnetization to leading non-trivial order in $h$,
\begin{multline}\label{eq:mag-scaling-classical}
    \chi_\perp^{-1} M^z(\bvec{r})  = h- \frac{ h^3}{4\pi^2 \rho_s^2 S^2} K_0 \Bigl(\frac{h}{cS} r  \Bigr)^2  \\
    - \frac{h^3}{2\pi \rho_s c^2S^2} J_A \sin(2\theta ) K_2 \Bigl(\frac{h}{cS} r  \Bigr),
\end{multline}
where $K_n(x)$ denote modified Bessel functions of second kind with the large-$x$ asymptotic $K_n(x) \sim \eu^{-x} \sqrt{\pi/(2x)}$ and $\sin 2 \theta \equiv 2 x y/r$ corresponds to the $d_{xy}$-wave form factor.
Crucially, this implies that $ [K_0 ( h r/c )]^2$ decays exponentially faster than $K_2 (hr/c)$ for large $r =|\bvec{r}|$, and thus the long-distance behaviour of the impurity-induced magnetization is dominantly governed by the second term in Eq.~\eqref{eq:mag-scaling-classical}.
The vacancy-induced magnetization texture with $d_{xy}$-symmetry transforms in the same IR $B_2$ as the staggered magnetization, allowing to \emph{image} the symmetry of the altermagnetic order parameter in real space.
Our low-energy field-theory arguments are independent of microscopic details and  readily extended to other altermagnets, e.g.~with $g$-wave symmetry, by appropriately replacing $2\partial_x \partial_y$ in $\mathcal{H}$ by the corresponding form-factor derivative operators.

\begin{figure}[tb]
    \centering
    \includegraphics[width=0.85\linewidth]{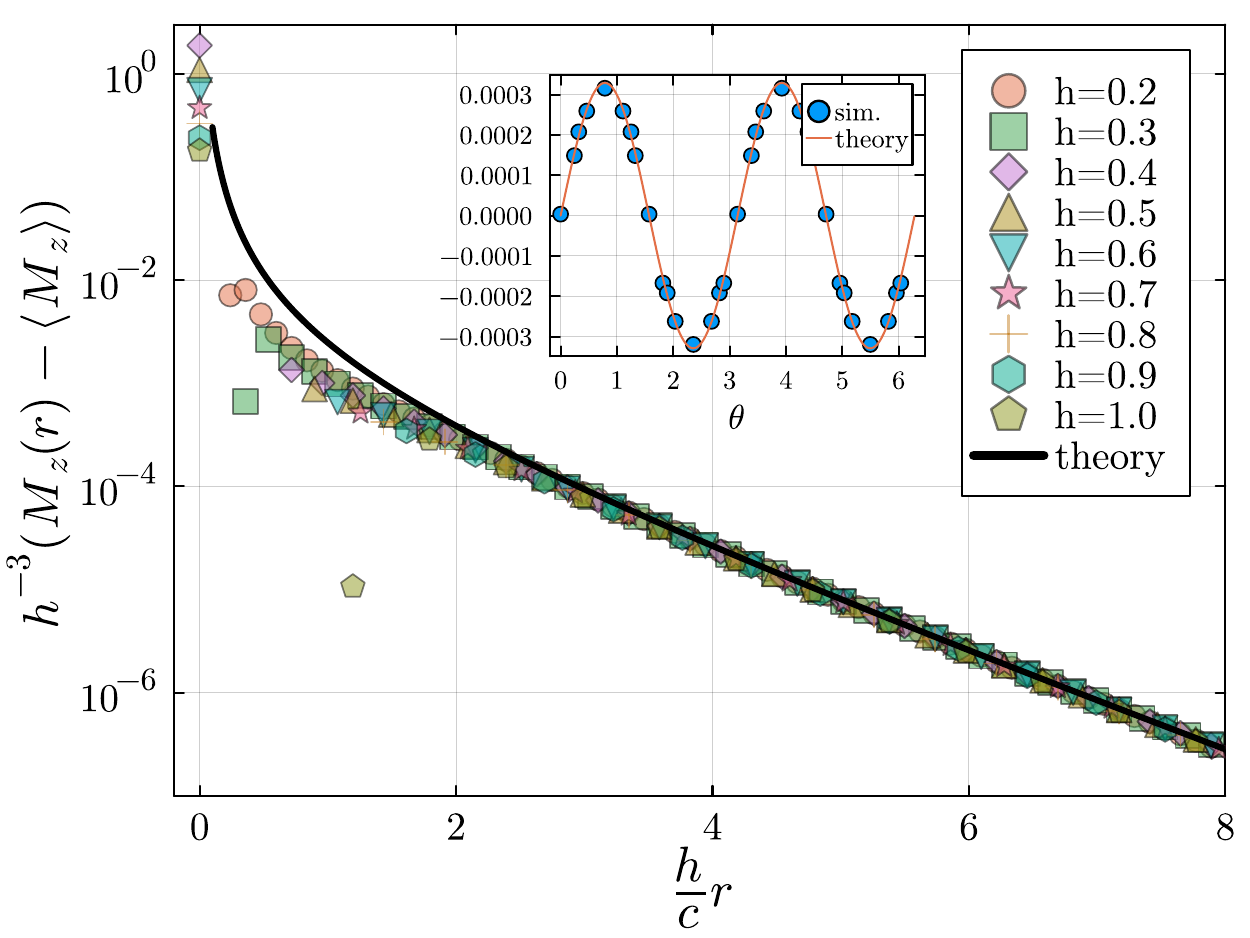}
    \caption{Scaling collapse of the magnetization profile 
    along the lattice diagonal, $\bvec{r}= r(1,1)$, compared to the theory prediction (solid line) from Eq.~\eqref{eq:mag-scaling-classical} for $J_2 = 0.3 J_1$ and $J_2' = 0$. In the inset, we consider $h=0.6 J_1$ and obtain $M_z(\bvec{r}(\theta))-1/8h$ as a function of distance $r$ away from the vacancy along various directions, labeled by $\theta$, and fit the large-$r$ asymptotics to the prefactor of
    $ K_2(hr/(Sc))$ and compare with the field-theory given in Eq.~\eqref{eq:mag-scaling-classical}.}
    \label{fig:scalingcollapse}
\end{figure}

We compare the universal field-theory result to our lattice-scale simulations by numerically evaluating $M^z(\bvec{r})$ along the diagonal $\bvec r = r(1,1)/\sqrt{2}$, for which $\sin 2\theta=1$ leads to a maximal effect.
The data for different field strengths exhibit a scaling collapse, see Fig.~\ref{fig:scalingcollapse}, and agree well with the continuum field-theory expression in Eq.~\eqref{eq:mag-scaling-classical}, which also reproduces the angular dependence of the texture.

\textit{Vacancy in quantum altermagnets at zero field.---}%
Turning to quantum models, the ordered moment $|\langle \bvec{S}_i \rangle| \leq S$ is renormalized by quantum fluctuations. 
Thus, a vacancy can lead to \emph{longitudinal} distortions of the collinear magnetic order even at zero field $\bvec{h} =0$, as illustrated in Fig.~\ref{fig:illustration-lattices}(c).

\begin{figure}
    \centering
    \includegraphics[width=0.99\linewidth]{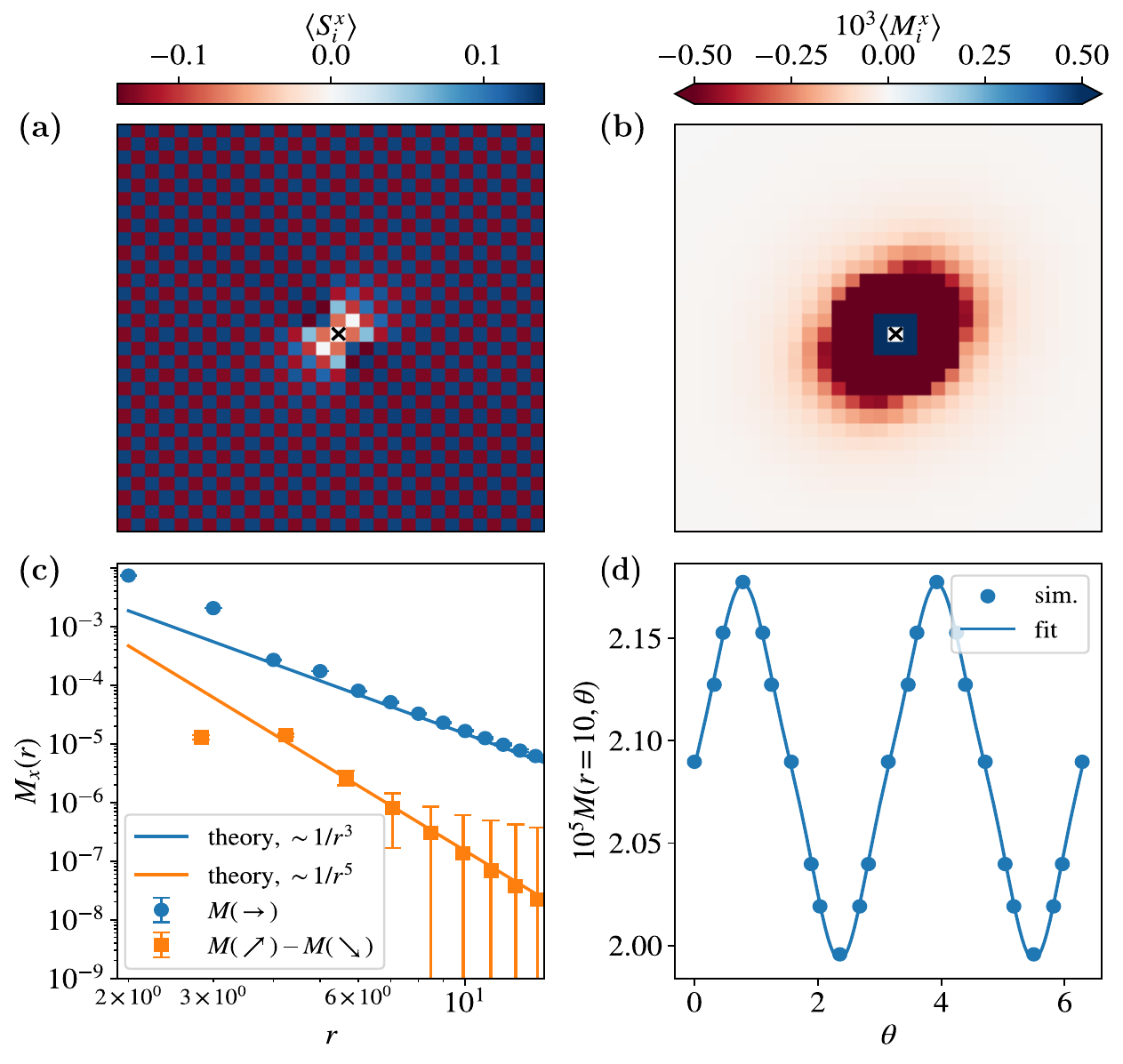}
    \caption{The quantum model for spins of length $S=1/2$ and $\bvec{h}=0$.
    (a)~The site dependent spin $\langle S^z_i \rangle$ evaluated in linear-spin wave theory for the checkerboard lattice with $J_2 = 0.6 J_1$. The position of the vacancy is marked with an 'x'.  (b) The local magnetization around the vacancy, again for $J_2 = 0.6 J_1$. We applied an additional Gaussian blur to $M_x(\mathbf{r})$ to reduce short distance artifacts close to the vacancy. (c) Decay of the magnetization in horizontal direction (blue) and the difference $M_x(r,r)-M_x(r,-r)$ (orange) as a function of distance from the vacancy for $J_2 = 0.1J_1$ compared to the theory in Eq.~\eqref{eq:m-x-quantum}. The errorbars indicate numerical uncertainties arising in finite-size scaling. (d) Angular dependence of the magnetization at a constant distance of the vacancy \cite{Footnote_fig_Altermagnet_panel} with $J_2 =0.4J_1$ and a fit of this data with function $a+ b\sin(2\theta)+c\sin(2\theta)^2+d\sin(2\theta)^3$, accounting for higher order anharmonic contributions beyond Eq.~\eqref{eq:m-x-quantum}. The fit parameters are $a = (2.1\pm 0.001) 10^{-5} , b = (6.5\pm0.2) 10^{-7} , c = (-2.1\pm 1.5) 10^{-8}$ and $d = (2.7\pm0.3) 10^{-7} $.}
    \label{fig:quantumAltermagnet_results}
\end{figure}

We demonstrate this explicitly for the checkerboard lattice Heisenberg model in Eq.~\eqref{eq:h-j}, but now with $S=1/2$ moments.
Corrections from quantum fluctuations in (semi-)classically ordered state can be obtained via spin-wave theory in a $1/S$ expansion, see End Matter for details.
The ordered moment on each site $\langle S^x_i \rangle$ for finite altermagnetic coupling $J_A = 0.6 J_1$, extrapolated to thermodynamic limit, is shown in Fig.~\ref{fig:quantumAltermagnet_results}(a).
The magnetic order acquires an anisotropic distortion in the vicinity of the vacancy which is particularly pronounced when comparing the $\hat{x} + \hat{y}$-axis with the $\hat{x}-\hat{y}$-axis.
As before, we obtain the locally averaged uniform magnetization $\bvec{M}(\bvec{r})$ using a Gaussian filter, shown in Fig.~\ref{fig:quantumAltermagnet_results}(b).

\textit{Nonlinear sigma model analysis.---}%
We complement our microscopic simulations with a field theory analysis.
For the bulk altermagnet, we obtain a Nonlinear Sigma Model (NLSM) for $\bvec{N} \equiv \bvec{N}(\bvec{r},\tau)$ from the coherent state path integral (Euclidean time), yielding \cite{NLSM_Altermagnet}
\begin{multline} \label{eq:NLSM-am}
        \mathcal{S}_\mathrm{bulk}[\bvec{N}] = \int d^2 \bvec{r} \int d \tau \;  \frac{\chi_\perp}{2}(\partial_\tau \bvec{N})^2 +  S^2 \frac{\rho_s}{2}(\nabla \bvec{N})^2
        \\
        +\iu S \chi_\perp J_A\bvec{N} \cdot (\partial_\tau \bvec{N} \times 2\partial_x \partial_y \bvec{N}).
\end{multline}
In writing $\mathcal{S}[\bvec{N}]$, we have integrated out the magnetization with the saddle-point configuration
\begin{equation} \label{eq:mag-quantum}
        \bvec{M} = \chi_\perp \bigl[\iu  \bvec{N} \times \partial_\tau \bvec{N}
    - S J_A 2\partial_x \partial_y \bvec{N}+ S J_A \bvec{N}(\bvec{N} \cdot2 \partial_x \partial_y \bvec{N}) \bigr],
\end{equation}
where the first term encodes the dynamical transverse magnetization conjugate to $\bvec{N}$, and the $J_A$-dependent terms reproduce the static altermagnetic contribution obtained previously in Eq.~\eqref{eq:mag-classical}.
Combined with the Berry phase term of the coherent state path integral, this contribution induces the altermagnetic term in Eq.~\eqref{eq:NLSM-am}, also discussed in Ref.~\onlinecite{NLSM_Altermagnet}, and we assume a cancellation of the remaining (staggered) Berry phase terms \cite{sachdev11}.

However, adding a vacancy, say at site $\bvec{r}= 0$, introduces an additional \emph{uncompensated} Berry phase \cite{sachdev99,sachdev03}, beyond the energetic modifications in Eq.~\eqref{eq:h-vac}.
This is accounted for by the effective vacancy action
\begin{align}
        \mathcal{S}_\text{vac}[\bvec{N}] =  &\int \du \tau \Big[ -\iu S(\bvec{A} \cdot \partial_\tau \bvec{N}) \big|_{\bvec{r}=0}  \notag
        \\
        &-\! \iu S g_A \bvec{N} \!\cdot \! (\partial_\tau \bvec{N} \!\times \! 2\partial_x \partial_y \bvec{N})\big|_{\bvec{r}=0} \!+\! f_{h=0}(\bvec{M},\bvec{N})\Big],
        \label{eq:NSLM_action_vacancy} 
\end{align}
where $\bvec{A}(\bvec{N})$ is the Berry connection encoding the geometry of order-parameter space, satisfying $\nabla_\bvec{N} \times \bvec{A}(\bvec{N}) = \bvec{N}$ and $g_A$ is an effective vacancy coupling with bare value $g_A = J_A \chi_\perp$.

We account for order-parameter fluctuations by expanding the staggered magnetization around a mean-field configuration, see Eq.~\eqref{eq:condensed_Neel_expansion}.
The resulting bulk action $\mathcal{S}_\mathrm{bulk}[n_y,n_z]$ then describes two modes with dispersion $\omega_{\pm}(\bvec{k}) \approx c |\bvec{k}| \pm S J_A k_x k_y$. Their linear dispersion (as $\bvec{k} \to 0$) is required by Goldstone's theorem, but note the momentum-dependent splitting (with a $d_{xy}$-wave form factor) characteristic for altermagnets \cite{maier23,liu24,NLSM_Altermagnet}.
Further, we similarly expand $\bvec{N}(\bvec{r},\tau)$ in the vacancy action $\mathcal{S}_\mathrm{vac}[\bvec{N}] \to \mathcal{S}_\mathrm{vac}[n^y,n^z]$ and in Eq.~\eqref{eq:mag-quantum}.
This allows us to relate the magnetization $\langle \bvec{M}(\bvec{r},\tau) \rangle$ to the response of the Gaussian fields $n^y$ and $n^z$ to vacancy-induced $(0+1)$-dim. defect terms.
Deep in the ordered phase, the transverse fluctuations $n^y$ and $n^z$ are small and we thus treat the defect term perturbatively (formally controlled by a large order parameter stiffness $\rho_s$ and $\chi_\perp J_A \ll 1$).
Carrying out this program (see \cite{suppmat} for details) and using the bare values for the vacancy couplings, we obtain the leading-order decay of the uniform magnetization profile as (recall that $\langle M^y \rangle = \langle M^z \rangle \equiv 0$ by symmetry)
\begin{align}
     \langle M^x (\bvec{r},\tau)\rangle  = &- \frac{1}{64 \pi\sqrt{\rho_s \chi_\perp} } \frac{1}{r^3}
     \notag
     \\ 
     &-  J_A\sin(2 \theta) \frac{645  \chi_\perp}{1024 \pi\sqrt{\rho_s \chi_\perp}} \frac{1}{ r^5}  +\ldots,\label{eq:m-x-quantum}
\end{align}
where the $\ldots$ are subleading terms including higher orders of $J_A$. Note that the $1/r^{3}$-decay directly follows from the linear dispersion of Goldstone modes and has been reported before for conventional antiferromagnets \cite{luescher05,metlitski07}.
Thus, in an altermagnet with $J_A \neq 0$, the vacancy induces a distortion with a -- to leading order -- $ \sin 2\theta$-angular dependence, which corresponds precisely to the $d_{xy}$-wave form factor associated with the $B_2$ IR of the altermagnetic order parameter.

We compare our field-theoretic predictions with LSWT simulations. As a consistency check, in Fig.~\ref{fig:quantumAltermagnet_results}(c), we first verify that the magnetization indeed exhibits a $r^{-3}$-decay along the $\hat{x}$-axis away from the vacancy, where \eqref{eq:m-x-quantum} suggests altermagnetic corrections to be small. Proceeding, we take the antisymmetric combination of $M^x(r,\theta=\pi/4) -M^x(r,\theta=-\pi/4)$ to isolate the altermagnetic contribution.
The data matches well \emph{both} the expected $r^{-5}$-power law of the altermagnetic contribution and its analytically obtained prefactor.
We further consider the angular dependence of the magnetization along a circle of constant radius $M^x(r_0,\theta)$ around the vacancy.
As shown in Fig.~\ref{fig:quantumAltermagnet_results}(d), the numerical data exhibit a slight deviation from the ideal $\sin(2\theta)$ profile predicted in Eq.~\eqref{eq:m-x-quantum}. This distortion is expected, since the neglected higher-order contributions contain terms of the form $J_A^n \sin(2\theta)^n$, which spoil the harmonic angular dependence. To quantify this effect, we fit the angular profile using the leading $\sin(2\theta)$ contribution supplemented by the expected higher harmonics.


\textit{Conclusion.---}%
We have shown that the magnetization texture around a single vacancy in a collinear altermagnet inherits the symmetry of the altermagnetic order parameter, both for classical moments and for quantum spins. We have demonstrated this explicitly for realizations of $d$-wave altermagnets and provided a general symmetry-based argument that applies to altermagnets more broadly.
Several directions for future work remain open: this includes the study of a finite \emph{density} of vacancies and effective interactions between the resulting impurity-induced textures. Beyond the static response considered here, it would also be worthwhile to study the dynamical response associated with quantum impurities.


\textit{Acknowledgements.---}%
We gratefully acknowledge discussions with S.~D.~Lundemo, A. Rosch, C.~Schrade, A.~Sudbø and M. Vojta.
This work is funded by the Deutsche Forschungsgemeinschaft (DFG, German Research Foundation) through SFB 1238, project ID 277146847 (RB and UFPS), and the Emmy Noether Program, project ID 544397233, SE 3196/2-1 (UFPS). R.B. acknowledges support from the Studienstiftung des deutschen Volkes.
MSS acknowledges funding by the European Union (ERC-2021-STG, Project 101040651— SuperCorr). Views and opinions expressed are however those of the authors only and do not necessarily reflect those of the European Union or the European Research Council Executive Agency. Neither the European Union nor the granting authority can be held responsible for them.


\bibliography{Altermagnetism}

\begin{thebibliography}{62}
\bibitem [{sup()}]{suppmat}%
  \BibitemOpen
  \href@noop {} {\bibinfo {title} {See the \uppercase{S}upplemental
  \uppercase{M}aterials for details on the classical field theory, nonlinear
  sigma model analysis, and a classical simulations of a honeycomb-lattice
  altermagnet, which also includes
  \uppercase{R}efs.~\onlinecite{Camerano_Honeycomb_altermagnet,sarkar25}.}}\BibitemShut
  {Stop}%
\bibitem [{Foo()}]{Footnote_fig_Altermagnet_panel}%
  \BibitemOpen
  \href@noop {} {}\bibinfo {note} {If no lattice site exists at the position
  $\bvec{r}(r,\theta)$, the magnetization at this point is obtained by
  interpolation from the neighboring lattice sites.}\BibitemShut {Stop}%
\end{thebibliography}


\clearpage
\appendix

\section{End Matter}

\subsection{Real-space spin-wave theory for vacancy-induced magnetization textures}

\begin{figure}[htb]
    \centering
    \includegraphics[width=1.0\linewidth]{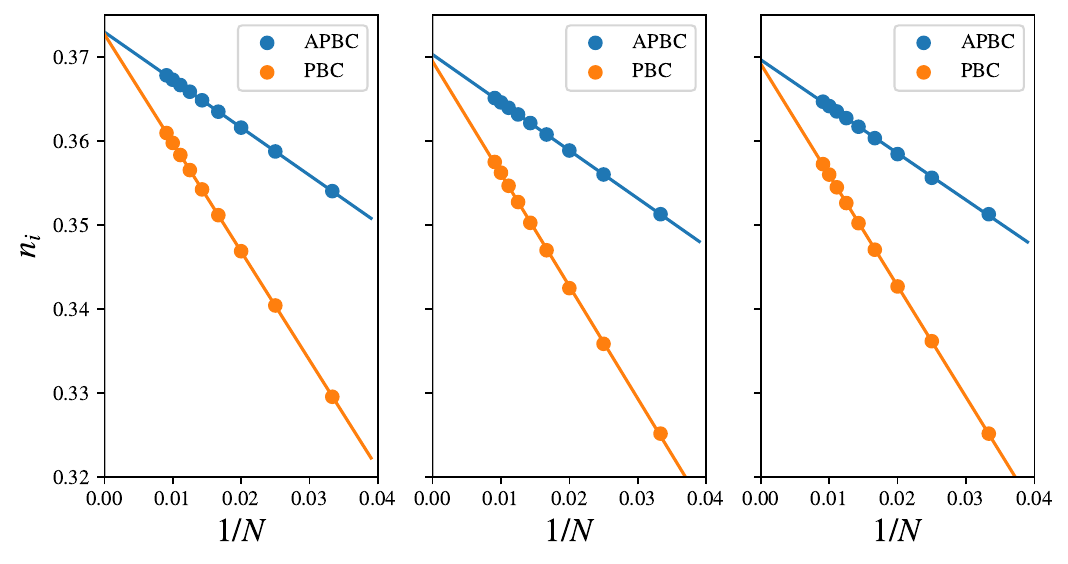}
    \caption{Extrapolation to infinite system size of the bosonic occupation number $n_i = \langle a_i^\dagger a_i \rangle$ on the checkerboard lattice. Shown are $n_i(1/L)$ at sites $\bvec{r}=(0,5), (0,10)$ and $(0,20)$, from left to right. We use system sizes $L \in \{30,40,...,110\}$, $J_2 = 0.6$ and compare imposing periodic and antiperiodic boundary conditions (along $\hat{x}$- and $\hat{y}$-axis) for the bosons.}
    \label{fig:lswt-extrapol}
\end{figure}

Quantum fluctuations in a state with (semi-)classical magnetic order can be systematically analyzed in a large-$S$ expansion.
To this end, we use the Holstein-Primakoff representation which, for a classical spin pointing along the $\hat{z}$-axis (local frame), reads $\tilde{S}^+ = \sqrt{2S-a^\dagger a} a$, $\tilde{S}^- = a^\dagger\sqrt{2S-a^\dagger a}$ and $\tilde{S}^z = S-a^\dagger a$, with $a,a^\dagger$ being bosonic annihilation and creation operators, respectively.
Without loss of generality, we choose antiferromagnetic order along the $\hat{x}$-axis such that the spin operators in the global (lab) frame are given as $S^x_i = (-1)^i \tilde{S}^z_i$, $S^y_i = (-1)^i \tilde{S}^x$ and $S^z_i = \tilde{S}^y_i$.
Inserting into the Eq.~\eqref{eq:h-j}, the Hamiltonian can then be organized in a $1/S$ expansion $H = S^2 H^{(0)} + S H^{(2)} + \mathcal{O}(S^0)$
where $H^{(0)}$ corresponds to the classical approximation and $H^{(2)}$ denotes the Hamiltonian which is quadratic in the bosonic operators $a_i,a_i^\dagger$ (note that no linear terms in $a_i$ occur if the chosen classical reference state is stable).
In the presence of a vacancy, $H^{(2)}$ is not translationally invariant, such that finding the eigenmodes necessitates a real-space Bogoliubov transformation.
Writing the bosonic Hamiltonian as $H^{(2)} = \frac{1}{2}  \underline{\psi}^\dagger \underline{\underline{H}} \underline{\psi}$ with $\underline{\underline{H}}$ denoting a $2N \times 2N$ matrix and the spinor $\underline{\psi} =(a_1,\dots, a_N, a_1^\dagger, \dots, a_N^\dagger)^\top$,
the eigenmodes of the Hamiltonian are obtained by a paraunitary transformation
\begin{equation} 
	\underline{\psi} = \begin{pmatrix} \underline{\underline{U}}  & \underline{\underline{V}}^\ast \\ \underline{\underline{V}} & \underline{\underline{U}}^\ast \end{pmatrix} \underline{\gamma},
\end{equation}
where $\underline{\underline{U}},\underline{\underline{V}}$ are $N \times N$ matrices and $\underline{\gamma} = (\gamma_1, \dots, \gamma_N)^\top$ is a spinor of eigenmodes with eigenenergy $\epsilon_\mu$, i.e. $H^{(2)} = \sum_\mu \epsilon_\mu \gamma_\mu^\dagger \gamma_\mu + \mathrm{const.}$
We construct $\underline{\underline{U}},\underline{\underline{V}}$ following the algorithm by Colpa \cite{colpa78}, see also Ref.~\onlinecite{wessel05} for an alternative approach.

Note that the presence of a vacancy at site $i=0$ implies that the Hamiltonian becomes independent of $\bvec{S}_{i=0}$, and correspondingly $a_{i=0}$ and $a_{i=0}^\dagger$ are zero modes of $H^{(2)}$, which we explicitly project out to avoid numerical instabilities.
The ordered moment on site $i$ (to order $(1/S)^0$) is then obtained as (at zero temperature $T=0$)
\begin{equation}
    \langle S^x_i \rangle = (-1)^i \left(S - \langle a_i^\dagger a_i \rangle_{(2)} \right),
\end{equation}
and $\langle a_i^\dagger a_i \rangle_{(2)} = \sum_{n=1}^N |V_{in}|^2$.
We consider finite-size grids of $L\times L$ lattice sites with $L \in \{30,40,...,110\}$.
To account for finite-size effects, we then extrapolate the data $\langle S_i^x \rangle(1/L)$ to $1/L = 0$.
While imposing periodic or antiperiodic boundary conditions for the Holstein-Primakoff bosons $a,a^\dagger$ yields the same results in thermodynamic limit, we find that the latter exhibit reduced finite-size effects and allow for significantly faster convergence, as demonstrated for a selection of three sites in Fig.~\ref{fig:lswt-extrapol}.


\end{document}



\title{Supplemental Material:\\
Anisotropic vacancy-induced magnetization textures in altermagnets}

\author{Ruben Burkard}
\affiliation{Institut für Theoretische Physik, Universit\"at Köln, Zülpicher Straße 77a, 50937 Köln, Germany}

\author{Mathias S. Scheurer}
\affiliation{Institute for Theoretical Physics III, University of Stuttgart, 70550 Stuttgart, Germany}

\author{Urban F. P. Seifert}
\affiliation{Institut für Theoretische Physik, Universit\"at Köln, Zülpicher Straße 77a, 50937 Köln, Germany}


\date{\today}

\maketitle

\tableofcontents


\section{Field theory for classical altermagnet}

We consider the classical Heisenberg model describing spins of length $\sqrt{\bvec{S}^2}= S$ on the checkerboard lattice with Hamiltonian $H = H_0+H_h$ and 
\begin{subequations}
\begin{align}\label{eq:Hamiltonian_AM_checkerboard}
    H_0
    &= J_1\sum_{\langle ij\rangle} \bvec{S}_i \cdot \bvec{S}_j  + J_2 \sum_{i }  \bvec{S}_i \cdot \bvec{S}_{i+\hat{x}+(-1)^i\hat{y}}+ J_2' \sum_{i } \bvec{S}_i \cdot \bvec{S}_{i+\hat{x} 
    -(-1)^i\hat{y}},
    \\
    H_h &= -\sum_i \bvec{h} \cdot \bvec{S}_i , \label{eq:Hamiltonian_AM_checkerboard_h}
\end{align}\label{TotalClassicalHamiltonian}
\end{subequations}
where we define the base vectors $\hat{x} = (1,0)$ and $\hat{y} = (0,1)$. In the next step, we decompose each spin vector $\bvec{S}$ into the uniform magnetization $\bvec{M}$ and staggered magnetization $\bvec{N}$ via
\begin{align}\label{eq:Spin_decomposition}
    \bvec{S}( \bvec{r}) = (-1)^\bvec{r} \bvec{N}(\bvec{r}) \sqrt{S^2-\bvec{M}^2(\bvec{r})}+\bvec{M}(\bvec{r}),
\end{align}
where we have to impose $\bvec{N}^2=1$ and $\bvec{N} \cdot \bvec{M} = 0$ to satisfy the length constraint of each spin.
Inserting Eq.~\eqref{eq:Spin_decomposition} into Eq.~\eqref{TotalClassicalHamiltonian} and performing a gradient expansion yields the Hamiltonian density
\begin{align}
    \mathcal{H}_\mathrm{bulk} =  \quad &\frac{1}{  2 \chi_\perp} \bvec{M}^2 + S^2\frac{\rho_s}{2} (\nabla \bvec{N})^2 + SJ_A \bvec{M} \cdot 2\partial_x \partial_y \bvec{N} - \bvec{h}\cdot\bvec{M}+\lambda \bvec{M} \cdot  \bvec{N}+ \mu (\bvec{N}^2-1) ,
\end{align}
with parameters $\chi_\perp= 1/(8J_1)$, $\rho_s=J_1-J_2-J_2'$ and $J_A= J_2-J_2'$. Here, $\mu$ and $\lambda$ are Lagrange multipliers, enforcing the aforementioned constraints. Adding a vacancy at site $\bvec{r}=0$ (without loss of generality) amounts to adding $H\to H + H_\mathrm{vac}$ with the ``vacancy Hamiltonian'',
\begin{align}\label{eq:Hamiltonian_Vacancy}
    H_\mathrm{vac}  = -J_1\Bigl( \sum_{ \delta \in \text{NN} } \bvec{S}_\bvec{0} \cdot \bvec{S}_\delta \Bigr)   -J_2 \bvec{S}_\bvec{0} \cdot \bvec{S}_{\hat{x}+\hat{y}}-J_2 \bvec{S}_\bvec{0} \cdot \bvec{S}_{-\hat{x}-\hat{y}}- J_2'  \bvec{S}_\bvec{0} \cdot \bvec{S}_{\hat{x}-\hat{y}}- J_2'  \bvec{S}_\bvec{0} \cdot \bvec{S}_{-\hat{x}+\hat{y}}+ \bvec{h} \cdot \bvec{S}_\bvec{0} ,
\end{align}
which leads via a gradient expansion to the Hamiltonian density
\begin{align} \label{eq:vac_terms_Ham}
    \mathcal{H}_\mathrm{vac} = 
    \delta(\bvec{r})\bigl[ \begin{aligned}[t]   
    &S\rho_s \bvec{M} \cdot \nabla^2 \bvec{N} -S^2J_A\bvec{N} 2\partial_x\partial_y \bvec{N} -\frac{1}{   \chi_\perp} \bvec{M}^2 + S^2\rho_s \bvec{N} \nabla^2  \bvec{N} 
    - SJ_A \bvec{M} \cdot 2\partial_x \partial_y \bvec{N} + \bvec{h}\cdot\bvec{M} +S\bvec{h} \cdot \bvec{N}\bigr]. \end{aligned}
\end{align}
The classical equations of motion for $\bvec{M}$ obtained from $\mathcal{H}_0+\mathcal{H}_\mathrm{vac}$ yield
\begin{align} \label{eq:classical_mag_formula}
    \bvec{M}(\bvec{r}) = \chi_\perp \left\{
    \begin{matrix}
        &\bvec{h}- \bvec{N} (\bvec{h}\cdot \bvec{N})-SJ_A 2\partial_x \partial_y \bvec{N}+SJ_A \bvec{N}(\bvec{N} \cdot 2 \partial_x \partial_y \bvec{N}  )  & \text{for } \bvec{r}\neq \bvec{0}
        \\
        &S \rho_s \nabla^2 \bvec{N}-S\rho_s \bvec{N}(\bvec{N} \cdot \nabla^2 \bvec{N})  & \text{for } \bvec{r}= \bvec{0}
    \end{matrix} \right. ,
\end{align}
which can be inserted back into the total Hamiltonian, which then depends only on $\bvec{N}$: 
\begin{align}
\mathcal{H}_\mathrm{bulk}+\mathcal{H}_\mathrm{vac} =  &\frac{1}{2} \chi_\perp \bigl(\bvec{h} \cdot \bvec{N} \bigr)^2 +S^2\frac{\rho_s}{2} (\nabla \bvec{N})^2 +\mu (\bvec{N}^2-1)  \notag\\ &+\delta(\bvec{r}) \Bigl[-\frac{1}{2} \chi_\perp \bigl(\bvec{h} \cdot \bvec{N} \bigr)^2  +S^2 \rho_s \bvec{N} \cdot \nabla^2 \bvec{N}-S\chi_\perp J_A \bvec{h} \cdot
 2\partial_x \partial_y \bvec{N}+S\bvec{h} \cdot \bvec{N}-
 S^2J_A\bvec{N}2\partial_x \partial_y \bvec{N} \Bigr].
\end{align}  
Note that we dropped a term proportional to the total derivative $\bvec{h} \cdot 2\partial_x \partial_y \bvec{N} $. 
Next, we assume that $\bvec{N}$ is condensed in $x$-direction, and we expand for small deviations from the homogeneous staggered magnetization $\bvec{N}=(1,0,0)^\top$, i.e.
\begin{align} \label{eq:condense_N}
    \bvec{N}  = \Bigl(\sqrt{1-n_y^2-n_z^2 }\ ,\  n_y \ ,\ n_z  \Bigr) , \ \ \text{with } n_{y,z} \ll1.
\end{align}
We choose $\bvec{h}=h \hat{z}$, which means that $n_y$ and $n_z$ are decoupled up to quadratic order \cite{eggert07}, and we can write down the field theory for the $n_z$ component with 
\begin{subequations}
\begin{align}
        \mathcal{H}_\mathrm{bulk}[n_z] &= 
   \frac{1}{2} \chi_\perp h^2  n_z ^2+S^2\frac{\rho_s}{2} (\nabla n_z)^2 , \\
    \mathcal{H}_\mathrm{vac} [n_z] &= \delta(\bvec{r}) \Bigl[ S h n_z-S\chi_\perp J_A h 
 2\partial_x \partial_y n_z -S^2J_A   n_z
 2\partial_x \partial_y n_z +S^2\rho_s n_z  \nabla^2 n_z \Bigr] +\mathcal{O}(h^2).
\end{align}
\end{subequations}
Note that a vacancy acts as a UV defect and thus, at low energies, the effective vacancy couplings will flow away from their bare (lattice-scale) values.
However, the leading long-distance behavior derived below is cutoff-independent and we may therefore continue with the bare coupling constants.
In order to get the spatial dependence of $n_z$, we define the generating functional
\begin{equation}
Z[J] = \int \mathcal{D}[n_z(\bvec r)] 
\exp\!\left[-\beta \int d^2 \bvec{r} \big( \mathcal{H}_\mathrm{bulk}[n_z(\bvec r)] + \mathcal{H}_\mathrm{vac}[n_z(\bvec{r})] \big) + \int d^2 \bvec{r} J(\bvec{r})  n_z(\bvec{r})\right],
\end{equation}
and do perturbation theory for small $\mathcal{H}_\text{vac}$, keeping terms up to order $J_A$ and $h^3$ to get
\begin{align}
    \langle n_z(\bvec{r})\rangle =  &\beta Sh \bigl\langle n_z(\bvec{r})  n_z(0) \bigr\rangle_0- \beta S\chi_\perp J_A h \bigl\langle n_z(\bvec{r}) 2 \partial_x \partial_yn_z(0) \bigr\rangle_0-\beta^2S^3 h J_A
    \bigl\langle n_z(\bvec{r})  n_z(0)  n_z(0)
 2\partial_x \partial_y n_z(0) \bigr\rangle_0  \notag \\
 &+\beta^2S^3 h \rho_s
    \bigl\langle n_z(\bvec{r})  n_z(0)  n_z(0)
 \nabla^2 n_z(0) \bigr\rangle_0,
\end{align}
where $\langle \ldots \rangle_0$ denotes the expectation value evaluated in the theory without a vacancy.
Using Wick's theorem, we can rewrite this expression as 
\begin{align} \label{eq:Classical_Neel_z_Greens_fct}
    \langle n_z(\bvec{r})\rangle =  &\beta ShG(\bvec{r})-\beta S\chi_\perp J_A h  2\partial_x \partial_y G(\bvec{r})- \beta^2 S^3h J_A
    \bigl[G(0) 2\partial_x \partial_y G(\bvec{r})+G(\bvec{r}) 2 \partial_x \partial_y G(0) \bigr]
    \notag \\
    &+ \beta^2 S^3h \rho_s
    \bigl[G(0) \nabla^2 G(\bvec{r})+G(\bvec{r}) \nabla^2 G(0) \bigr],
\end{align}
with the free Green's function
\begin{equation}
    G(\bvec{r}) = \bigl\langle n_z(\bvec{r})  n_z(0) \bigr\rangle_0 =\frac{1}{\beta} \Bigl(\frac{1}{2} \chi_\perp h^2  -S^2\frac{\rho_s}{2} \nabla^2  \Bigr)^{-1} = \frac{1}{2\pi  \rho_s S^2 \beta} K_0 \Bigl(\frac{h}{cS} r  \Bigr) , 
\end{equation}
where $K_n$ are the modified Bessel functions of second kind and we define $c= \sqrt{\rho_s/\chi_\perp}$. Thus
we get
\begin{align} \label{eq:n_z_classical}
    \langle n_z(\bvec{r})\rangle =\frac{h}{2\pi \rho_sS} K_0 \Bigl(\frac{h}{cS} r  \Bigr)
    +\alpha h^3 K_0 \Bigl(\frac{h}{cS} r  \Bigr)
     +\gamma h^3 J_A  \sin(2 \theta)K_2 \Bigl(\frac{h}{cS} r  \Bigr),
\end{align}
where $\alpha$ and $\gamma$ are parameters that depend on a UV cutoff, which emerge from vacancy-induced ``contact terms'' of the long-wavelength fields such as $\langle n_z(0) n_z(0) \rangle $.
We stress that Eq.~\eqref{eq:n_z_classical} is independent of temperature and thus in particular also applies at zero temperature $T=\beta^{-1} = 0$.
Note that the first term in Eq.~\eqref{eq:n_z_classical} has previously been derived for conventional antiferromagnets in Ref.~\onlinecite{eggert07}.
Next, we use Eq.~\eqref{eq:classical_mag_formula} to calculate the magnetization at large distances from the vacancy and small $h$, giving
\begin{align}
    \chi_\perp^{-1}M^z   = h- \frac{ h^3}{4\pi^2 \rho_s^2S^2} K_0 \Bigl(\frac{h}{cS} r_0  \Bigr)^2
    - \frac{h^3}{2\pi \rho_s c^2S^2} J_A \sin(2\theta ) K_2 \Bigl(\frac{h}{cS} r_0  \Bigr) +\mathcal{O}(h^4).
\end{align}
This expression is independent of the UV cutoff up to order $\mathcal{O}(h^3)$.


\section{Quantum nonlinear sigma model analysis}%
We consider quantum spins at zero temperature and of length $S$ on the square lattice without external magnetic field. The Hamiltonian of the system is given by $H_0$ from Eq.~\eqref{eq:Hamiltonian_AM_checkerboard}.
We do a gradient expansion of the coherent state path integral of the partition function in Euclidean time \cite{Sachdev_2011}
\begin{align}
     Z = \int \mathcal{D}\bvec{N}  \mathcal{D}\bvec{M} \;   \; e^{-S_b[\bvec{N},\bvec{M}]}\; \; e^{-S_\text{H}[\bvec{N},\bvec{M}]},
\end{align}
where we decompose the spin vector into uniform magnetization and staggered magnetization using Eq.~\eqref{eq:Spin_decomposition}.
The action is given by $S_b+S_\text{H}$ with
\begin{subequations}
\begin{align}
    S_b[\bvec{N},\bvec{M}]& = \int d^2 \bvec{r} \int d \tau  \; iS \bvec{A} \cdot \partial_\tau \bvec{N}-i \bvec{M} \cdot (\bvec{N} \times \partial_\tau \bvec{N}) \\ 
    S_\text{H}[\bvec{N},\bvec{M}]& = \int d^2 \bvec{r} \int d \tau \; \frac{1}{2\chi_\perp} \bvec{M}^2 +  S^2 \frac{\rho_s}{2}(\nabla \bvec{N})^2 + S J_A\bvec{M} \cdot 2\partial_x \partial_y \bvec{N} + \lambda \bvec{M} \cdot \bvec{N} +\mu (\bvec{N}^2-1), \label{eq:quantum_NLSM_action_NM}
\end{align}
\end{subequations}
where $\lambda$ and $\mu$ are Lagrange multipliers that enforce $\bvec{N}\cdot \bvec{M} =0$ and $\bvec{N}^2=1$, ensuring that $\bvec{S}_i^2 = S^2$. We defined the ``Dirac monopole'' function $\bvec{A}(\bvec{N})$, satisfying $\nabla_\bvec{N} \times \bvec{A}(\bvec{N}) = \bvec{N}$.
Here, the action $S_b[\bvec{N},\bvec{M}]$ accounts for the accumulated Berry phase of the spins during imaginary time evolution, and the first term is expected to cancel out in two dimensions without a vacancy.
We then solve for the magnetization, obtaining
\begin{equation}
    \bvec{M} = \chi_\perp \bigl[i  \bvec{N} \times \partial_\tau \bvec{N}
    - S J_A 2\partial_x \partial_y \bvec{N}+ S J_A \bvec{N}(\bvec{N} \cdot 2\partial_x \partial_y \bvec{N}) \bigr].
\end{equation}
This allows us to integrate out $\bvec{M}$, yielding \cite{NLSM_Altermagnet}
\begin{align}
    S_b[\bvec{N}]+S_\text{H}[\bvec{N}]& = \int d^2 \bvec{r} \int d \tau \;  \frac{\chi_\perp}{2}(\partial_\tau \bvec{N})^2 +  S^2 \frac{\rho_s}{2}(\nabla \bvec{N})^2 +i S \chi_\perp J_A\bvec{N} \cdot (\partial_\tau \bvec{N} \times 2\partial_x \partial_y \bvec{N}).
\end{align}
Adding a vacancy at $\bvec{r}= 0$ will generate an uncompensated Berry phase in addition to the contributions discussed in Eq.~\eqref{eq:vac_terms_Ham}. This yields the contribution to the action
\begin{equation}
        S_\text{vac}[\bvec{N}] =  \int d^2 \bvec{r} \int d \tau  \; \delta(\bvec{r})\Bigl[ -i S(\bvec{A} \cdot \partial_\tau \bvec{N}) -i S \chi_\perp J_A\bvec{N} \cdot (\partial_\tau \bvec{N} \times 2\partial_x \partial_y \bvec{N})   - \frac{\chi_\perp}{2}(\partial_\tau \bvec{N})^2 +S^2 \rho_s \bvec{N} \cdot \nabla^2 \bvec{N}-
 S^2J_A\bvec{N}2\partial_x \partial_y \bvec{N} \Bigl].
\end{equation}
We again condense $\bvec{N}$ along the $x$-axis as in Eq.~\eqref{eq:condense_N}, but this time we find it convenient to use the complex representation
\begin{align}
    \bvec{N} = \Bigl(1-|\psi|^2 \, , \;\frac{\psi+\psi^*}{2} \sqrt{2-|\psi|^2}\;,\;\frac{\psi-\psi^*}{2i} \sqrt{2-|\psi|^2}\; \Bigr), \ \ \text{with } |\psi|^2\ll1.
\end{align}
Up to quadratic order in $|\psi|$ this gives
\begin{align} \label{eq:bulk_action_nlsm}
    S_b[\psi,\psi^*]+S_H[\psi,\psi^*]& = \int d^2 \bvec{r} \int d \tau \;  \chi_\perp |\partial_\tau \psi|^2 +S^2\rho_s |\nabla \psi|^2+ S \chi_\perp J_A\bigl[(\partial_\tau \psi^* )(2\partial_x \partial_y \psi) -  (\partial_\tau \psi ) (2\partial_x \partial_y \psi^*)\bigr], \\ 
    S_\text{vac}[\psi,\psi^*]& = \int d^2 \bvec{r} \int d \tau  \; \delta(\bvec{r})\Bigl[ \begin{aligned}[t]  &- \chi_\perp |\partial_\tau \psi|^2 -2S^2\rho_s |\nabla \psi|^2- S \chi_\perp J_A\bigl[(\partial_\tau \psi^* )(2\partial_x \partial_y \psi) -  (\partial_\tau \psi ) (2\partial_x \partial_y \psi^*)\bigr] \\&-\frac{S}{2} (\psi^* \partial_\tau \psi- \psi \partial_\tau \psi^*)+2J_A[(\partial_x\psi)(\partial_y\psi^*)+(\partial_x\psi^*)(\partial_y\psi)] \Bigr]  \end{aligned},
\end{align} 
where we chose the gauge
\begin{align}
    \bvec{A}(\bvec{N}) = \frac{1}{|\bvec{N}|(|\bvec{N}|+N_x)} (0,-N_z, N_y).
\end{align}
The magnetization in $x$-direction takes the form
\begin{align}
    M^x &= \chi_\perp \Bigl[\psi^* \partial_\tau \psi - \psi \partial_\tau \psi^* +SJ_A (\psi^* 2\partial_x \partial_y \psi+\psi 2\partial_x \partial_y \psi^*) \Bigr]  +\mathcal{O}(|\psi|^3).
\end{align}
We can now do perturbation theory for small $ S_\text{vac}$ and get up to order $J_A$
\begin{align}
    \langle M^x (\bvec{r})\rangle = &-\chi_\perp\frac{S}{2} \int d\tau  \; \Bigl\langle  \bigl(\psi^* \partial_\tau \psi - \psi \partial_\tau \psi^*\bigr) \Big|_{(\bvec{r},0)} \;  \bigl(\psi^* \partial_\tau \psi-\psi \partial_\tau \psi^* \bigr)\Big|_{(0,\tau)}\Bigr\rangle _0
    \notag\\ 
    &- S\chi_\perp^2 J_A\int d\tau\Bigl\langle  \bigl(\psi^* \partial_\tau \psi - \psi \partial_\tau \psi^*\bigr) \Big|_{(\bvec{r},0)} \;  \bigl[ (\partial_\tau \psi^* )(2\partial_x \partial_y \psi)- (\partial_\tau \psi )(2\partial_x \partial_y \psi^*) 
    \bigr]\Big|_{(0,\tau)}\Bigr\rangle _0
        \notag\\ 
    &- S\chi_\perp^2 J_A\int d\tau\Bigl\langle  \bigl(\psi^* 2\partial_x \partial_y \psi+\psi 2\partial_x \partial_y \psi^*\bigr) \Big|_{(\bvec{r},0)} \;  (\partial_\tau \psi )(\partial_\tau \psi^*)\Big|_{(0,\tau)}\Bigr\rangle _0
\notag\\ 
    &- 2S^3\chi_\perp \rho_s J_A\int d\tau\Bigl\langle  \bigl(\psi^* 2\partial_x \partial_y \psi+\psi 2\partial_x \partial_y \psi^*\bigr) \Big|_{(\bvec{r},0)} \;  \bigl[ (\partial_x\psi)(\partial_y\psi^*)+(\partial_x\psi^*)(\partial_y\psi)
    \bigr]\Big|_{(0,\tau)}\Bigr\rangle _0
    \\
    &=  \int d\tau  \; \Bigl[ \begin{aligned}[t]
    &-2S\chi_\perp
     G(\bvec{r},\tau) \partial^2_\tau G(\bvec{r},-\tau) +2 S\chi_\perp^2 J_A G(\bvec{r},\tau)2\partial_x \partial_y \partial^2_\tau G(\bvec{r},-\tau)
     \\
     &- 4 S^3\chi_\perp \rho_s J_A \partial_x G(\bvec{r},\tau)2\partial_x^2 \partial_y G(\bvec{r},-\tau)
     - 4 S^3\chi_\perp \rho_s J_A \partial_y G(\bvec{r},\tau)2\partial_x \partial_y^2G(\bvec{r},-\tau)\Bigr], 
     \end{aligned}
\end{align}
where we have used imaginary time translational invariance and defined the field propagator
\begin{equation}
    G(\bvec{r},\tau) := \bigl\langle \psi(\bvec{r},\tau) \psi^* (0,0)\bigr\rangle_0.
\end{equation}
We first compute the propagator in momentum-frequency space using Eq.~\eqref{eq:bulk_action_nlsm}. Fourier transforming back to real space and imaginary time then gives the desired propagator, whose resulting integral representation is expanded in $J_A$. This gives
\begin{align}
    G(\bvec{r},\tau)  = \frac{1}{4 \pi \chi_\perp Sc }\frac{1}{\sqrt{S^2c^2\tau^2+r^2}}- \frac{3}{8 \pi \chi_\perp c}  J_A  \sin(2 \theta ) \frac{r^2 \tau }{\sqrt{S^2c^2\tau^2+r^2}^5}+\mathcal{O}(J_A^2),
\end{align}
where $\theta$ is the polar angle around the vacancy at $\bvec{r}=0$. Thus we get for the magnetization profile
\begin{align}
     \langle M^x (\bvec{r})\rangle  = \frac{1}{64 \pi\sqrt{\rho_s \chi_\perp} } \frac{1}{r^3}-  J_A\sin(2 \theta) \frac{645  \chi_\perp}{1024 \pi\sqrt{\rho_s \chi_\perp}} \frac{1}{r^5}    + \mathcal{O}(J_A^2) ,
\end{align}
which yields Eq.~(12) in the main text.


\section{Altermagnet on the honeycomb lattice}%
\begin{figure}
    \centering    \includegraphics[width=0.5\linewidth]{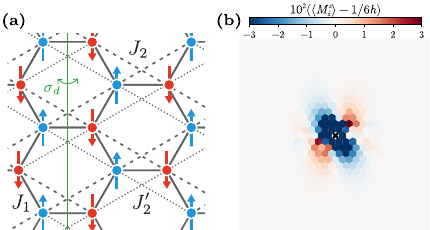}
    \caption{(a) Illustration of honeycomb lattice model, where $J_2 \neq J_2'$ results in inversion-symmetry breaking. As a result, the system becomes an altermagnet, protected by the combination $\Theta \sigma_d$ of time-reversal symmetry $\Theta$ and bond-center mirror operation $\sigma_d$. (b) The local magnetization around the vacancy, which is marked with 'x' for the parameters $h=0.8, J_2 = 0.5$ and $J_2'=-0.2$. To obtain a complete tiling of the two-dimensional plane, the magnetization at the center of each honeycomb plaquette is assigned the average magnetization of the surrounding sites.}
    \label{fig:AM_honeycomb_mag_profile}
\end{figure}

We turn to another realization of a $d$-wave altermagnet on the honeycomb lattice.
We consider classical spins with Hamiltonian $H = H_0+H_h$, where $H_h$ is defined as in Eq.~\eqref{eq:Hamiltonian_AM_checkerboard_h} and the Heisenberg exchange couplings are given by \cite{Camerano_Honeycomb_altermagnet,sarkar25}
\begin{align}\label{eq:Hamiltonian_AM_honeycomb}
    H_0
    &= J_1\sum_{\langle ij\rangle} \bvec{S}_i \cdot \bvec{S}_j  + \sum_{i\in A} \Bigl( J_2 \bvec{S}_i \cdot \bvec{S}_{i+\delta_2}+ J_2' \bvec{S}_i \cdot \bvec{S}_{i+\delta_2'}\Bigr)+
    \sum_{i\in B} \Bigl( J_2' \bvec{S}_i \cdot \bvec{S}_{i+\delta_2}+ J_2 \bvec{S}_i \cdot \bvec{S}_{i+\delta_2'}\Bigr),
\end{align}
with $\delta_2 = (3/2,\sqrt3/2 )^\top$ and $\delta_2' = (3/2,-\sqrt3/2 )^\top$, see Fig.~\ref{fig:AM_honeycomb_mag_profile}(a). Here, $i\in A(B)$ denotes that site $i$ belongs to sublattice $A(B)$. Note that for $J_2 \neq J_2'$, the sublattices are related by neither inversion nor translation symmetry. Crucially, the system retains a mirror symmetry $\sigma_d$ across an axis through bond centers, see also Fig.~\ref{fig:AM_honeycomb_mag_profile}(a).
In the presence of collinear magnetic order (antiparallel spins on neighboring sites), the combination $\Theta \sigma_d$ of the mirror operation $\sigma_d$ and time reversal $\Theta$ is a symmetry, thus rendering this system an altermagnet. For finite $J_2 \neq J_2'$, the system's point group is reduced from $C_{6v} \to C_{1v}$, with the staggered magnetization transforming in the (only) non-trivial IR of $C_{1v}$.
The magnetization profile around a single vacancy can be seen in Fig.~\ref{fig:AM_honeycomb_mag_profile}(b).

\bibliography{Altermagnetism}

\bibliographystyle{apsrev4-2}


\appendix